# Determinants of Women's Attitude towards Intimate Partner Violence: Evidence from Bangladesh

Md Tareq Ferdous Khan[1,2] & Lianfen Qian[3]

## Abstract

**Purpose:** The purpose of this study is to identify the important determinants responsible for the variation in women's attitude towards intimate partner violence (IPV). **Methods:** A nationally representative Bangladesh Demographic and Health Survey 2014 data of 17,863 women is used to address the research questions. In the study, two response variables are constructed from the five attitude questions, and a series of individual and community-level predictors are tested. The preliminary statistical methods employed in the study include univariate and bivariate distributions, while the adopted statistical models include binary logistic, ordinal logistic, mixed-effects multilevel logistic models for each response variable, and finally, the generalized ordinal logistic regression. **Results:** Statistical analyses reveal that among the individual-level independent variables age at first marriage, respondent's education, decision score, religion, NGO membership, access to information, husband's education, normalized wealth score, and division indicator have significant effects on the women's attitude towards IPV. Among the three community-level variables, only the mean decision score is found significant in lowering the likelihood. **Conclusions:** It is evident that other than religion, NGO membership, and division indicator, the higher the value of the variable, the lower the likelihood of justifying IPV. However, being a Muslim, NGO member, and resident of other divisions, women are found more tolerant of IPV from their respective counterparts. These findings suggest the government, policymakers, practitioners, academicians, and all other stakeholders to work on the significant determinants to divert women's wrong attitude towards IPV, and thus help to take away this deep-rooted problem from society.

## Background

Intimate partner violence (IPV) is considered a significant public health problem, and worldwide, almost one-third of women experience either physical or sexual violence (WHO, 2013). The consequences of IPV are well studied broadly either in terms of mental health or physical health problems to the victims (Campbell & Lewandowski, 1997; Campbell & Soeken, 1999; Golding,

[1] Division of Biostatistics and Bioinformatics, Department of Environmental and Public Health Sciences, College of Medicine, University of Cincinnati, Cincinnati, Ohio, USA
[2] Department of Statistics, Jahangirnagar University, Savar, Dhaka, Bangladesh
[3] Former Professor of Mathematical Sciences, Florida Atlantic University, Boca Raton, Florida, USA

1999; Cascardi, O'Leary, & Schlee, 1999; Coker et al. 2000, Campbell, 2002; Campbell et al., 2002, Bonomi et al. 2006; Sarkar, 2008; Vives-Cases, Ruiz-Cantero, Escribà-AgüirJuan, & Miralles, 2011; WHO, 2013; Devries et al. 2013; Okafor et al., 2018; Bacchus, Ranganathan, Watts, & Devries, 2018). The mental health problems are mainly in the form of depression, post-traumatic stress disorder, insomnia, substance abuse, and suicide attempt. The typical physical problems are physical injury, chronic pain, different kinds of headaches, gynecologic problems, sexually transmitted diseases, etc. The negative consequences of the IPV on pregnant women have demonstrated in Rodriguez et al. (2008) and Do et al. (2019). It is also well-documented that there is a significant effect on the children from the exposure of IPV (Campbell & Lewandowski, 1997; Bair-Merritt, Blackstone, & Feudtner, 2006; Hossain, Udo & Phillips, 2018).

Due to the severe health consequences of IPV, it has received huge attention from both research and policy perspectives all over the world. It is very essential to know how the victims justify the violence by their partners against themselves. If women are permissive to some form of violence, the perpetrators may take this opportunity to defend their offenses. A bunch of research has been carried out to identify the significant determinants of women's permissive attitude towards the violence in different countries. In this exertion, some authors have focused their concentration only on individual-level factors, while others put their efforts on identifying the community-level factors after adjusting the individual-level factors.

Among the individual-level factors, many research studies have shown that woman's educational status, sociodemographic status, autonomy or decision-making ability, wealth status, access to information and empowerment influence the attitude towards IPV (Lawoko, 2006; Lawoko, 2008; Uthman, Lawoko, & Moradi, 2009; Okenwa-Emegwa, Lawoko, & Jansson, 2016; Tran, Nguyen, & Fisher, 2016). Using the data of Kenya and Zambia, Lawoko (2008) showed that with the increase of education, women are more likely to reject IPV in Kenya, but it is not true in Zambia. For Kenya, they found the odds ratios of 0.86, 0.44, 0.31 for primary, secondary, and post-secondary compared to no education, while in Zambia, they found 2.97, 2.67, and 1.37 respectively. Wang (2016) in her review concluded that among all the factors associated with the attitude towards IPV, education is the most influencing factors, and other factors including wealth status, participation in the household decision, having access to information, traditional gender role followed by the households are subject to be varied due to different educational status.

Sayem et al. (2012) first studied women's attitude towards IPV by using 2010 survey data on 331 women selected from five disadvantaged areas in the Dhaka city, the capital of Bangladesh. In their study, they reported that women's attitude is affected by their different characteristics including education, age, age at marriage, micro-credit membership, the experience of any violence in the last 12 months preceding the survey, and exposure to media. The educational status of the husband is also evident as a significant determinant of the attitude. Jesmin (2017) used the nationally representative BDHS 2011 data to identify the social determinants that influence women's attitude in accepting IPV with a focus on community-level characteristics after controlling individual-level factors. It found that there is a significant negative correlation between the attitude towards IPV and two community-level variables, including the proportion of women living in poor households and the proportion of women illiterate, as well as the positive association between the attitude towards IPV and the community-level mean wife-beating score.

Khan and Islam (2018) demonstrated the impact of attitude towards IPV on the participation of the reproductive health services based on BDHS 2011 and 2014 data sets. They reported that strong rejection of IPV is positively associated with using modern contraception methods, seeking antenatal and postnatal care, receiving health care services from skilled health professionals, selecting health care services for delivery rather than home. Several other studies have been conducted in Bangladesh to show either the consequences of IPV or the factors associated with it or both along with the prevalence (Silverman, Gupta, Decker, Kapur & Raj, 2007; Åsling-Monemi, Naved, & Persson, 2008; Rahman, Hoque & Makinoda, 2011; Ziaei, Naved, & Ekström, 2012; Amin, Khan, Rahman & Naved, 2013; Hossain, Sumi, Haque, & Bari, 2014; Azzaiz-Baumgartner et al., 2014; Chowdhury et al., 2018, Ferdos et al., 2018, Parvin et al., 2018, De & Murshid, 2018), but not on the attitude towards IPV and its key factors.

This paper uses the Bangladesh Demographic and Health Survey 2014 (BDHS 2014) data (NIPORT & ICF, 2014) of 17,863 women to identify the determinants responsible for the variation in the attitude of the women towards IPV and provide the policy recommendations based on the significant determinants. In identification, both individual-level and community-level variables are examined by two response variables.

**Data and Methodology**

**Data**

The Bangladesh Demographic and Health Survey 2014 (BDHS 2014) data set of ever-married women aged 15 to 49 is used in this paper. The data is obtained from the ICF international (NIPORT & ICF Data, 2014). The BDHS 2014 is a nationally representative survey that is based on a two-stage stratified sampling. In the first stage of sampling, 600 enumeration areas (EA), each with an average of 120 households, were selected from a total of 293,579 EAs by probability proportional to size (PPS) sampling. Among these EAs, 207 were chosen from urban areas, and the rest 393 were from rural areas. In the second stage, a systematic sample was used to select an average of 30 households from each EAs. Overall, through this sampling design, a total of 18,000 households were selected, which eventually lead to complete interviews of 18,000 ever-married women. The valid data size is 17,863, ever-married women for analysis, and the rest of the women cannot be included because of non-response (NIPORT & ICF Report, 2014). The selected sample is hierarchical in structure as the sampled women are nested into 600 EAs, which are defined as the primary sampling units. Thus, the data has two levels, including individual women as the first level (individual-level hereafter) and primary sampling unit as the second level (community-level hereafter).

**Response variables**

Five questions to characterize the attitude of women towards IPV are whether the hitting or beating by husband is justified due to the situations (NIPORT & ICF Report, 2014) that

    (i)   if she goes out without telling him,
    (ii)  if she neglects the children,
    (iii) if she argues with him,
    (iv) if she refuses to have sex with him,
    (v)  if she burns the food,

The responses were recorded as either 'yes', 'no',  or 'don't know'.

Two response variables are constructed from the responses on the five questions. One of them is binary, while the other one is ordinal. The binary variable is taken 1 if the responses were 'yes' in at least any of the mentioned five situations and 0 if none of the answers were 'yes'. Another response variable is generated, having values ranging from 0 to 5, where the value is defined as the number of 'yes'  out of questions (i)-(v).

**Independent variables**

Both individual-level and community-level variables are considered to find the important determinants that may control the women's permissiveness attitude towards IPV. Among the individual-level variables, some are entirely associated with respondents; some are the characteristics of households, only one related to the husband. Three community-level variables are the proportion of women illiterate, the proportion living in the poorest household, which is defined as the bottom 20 percent based on wealth index and the mean decision score in the community.

According to the scale of measurements, some are continuous, and some are categorical. The characteristics of the respondents measured in ratio scale include age, age at first marriage, total years of education, autonomy or decision score, and access to information index. The decision score is computed from the participation of the respondent in decision making on the respondent's health care, major household purchases, and visits to family or relatives. Each of these variables is assigned 1 if the respondent alone or with husband makes the decision, otherwise 0 if someone else makes the decision. Then the summative rating scale is used to generate the decision score having the values 0 to 3. The access to information index is also calculated by using the summative rating scale from the three variables representing the frequency of watching television, listening radio, and reading the newspaper. The possible value for each variable is 0 if the answer was not at all, 1 if less than once a week and 2 if at least once a week. To characterize the husbands, their total years of education is considered as an independent variable. The binary variables which can describe the respondents include whether they engaged in any work in the last 12 months, their religion, whether they are the member of any non-government organization (NGO).

The household characteristics include the size of the household, place of residence either in rural or urban areas, division and normalized wealth score. In the 2014 survey time, Bangladesh was administratively divided into 7 divisions, and a dummy variable of Dhaka versus other divisions is used as an independent variable. Dhaka city, the capital of Bangladesh, is the divisional headquarter of the Dhaka division. Due to having different facilities, especially job opportunities, almost one-third of the population of Bangladesh is living in the Dhaka division. The wealth score is obtained from the asset variables by using the principal component analysis (NIPORT and ICF Report, 2014), and finally, the normalized score by classic min-max normalization approach is used in this paper.

**Statistical methods**

Statistical methods employed in this paper are categorized into two parts, including Preliminary Analysis and Statistical Models.

***Preliminary data analyses***

The frequency distributions of the response variables and the responses on the original questions are used to estimate the degree at which the women justify the violence against themselves by their husbands. Then, the bivariate table of the independent variables against the binary response variable is constructed to get the preliminary idea of the characteristics by which the attitude may vary. Statistical *t*-test and chi-square test are employed depending on the scale of measurement of the independent variables.

***Statistical models***

The statistical models listed in the following flowchart over the response variables are used to identify the important determinants that affect the women's attitude towards IPV. Among the first six models, the mathematical specification of the Model 6 is presented below as the models 1-5 can be derived as the special case of this model.

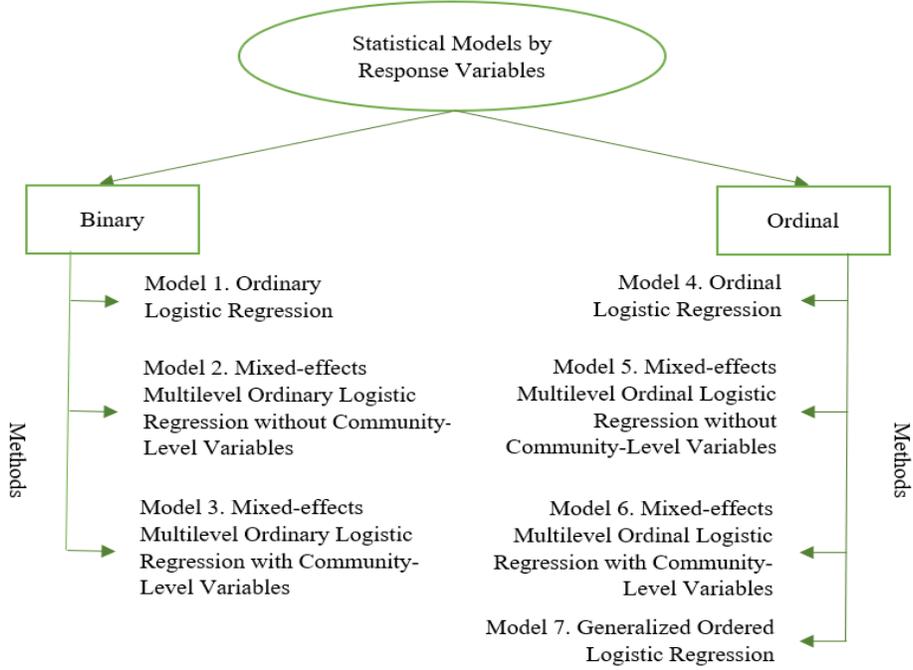

Figure 1: Flowchart of Statistical Models Fitted by the Response Variables

Let the latent variable for the ordered response variable is denoted by $y_{ij}^*$ and is defined by the model (Snijders & Bosker, 1999)

$$y_{ij}^* = \alpha + X_{ij}'\beta + \nu_j + Z_j'\theta + \varepsilon_{ij}, \; i = 1, 2, \cdots, n_j, \; j = 1, 2, \cdots 600, \tag{1}$$

where, $\alpha$ is the overall mean, $\nu_j$ is the random effect due to EA which makes intercept random, $X_{ij}$ is the vector of the individual-level independent variable corresponding to $i^{th}$ woman and $j^{th}$ EA, $\beta$ is the vector of individual-level regression coefficients, $Z_j$ is the vector of community-level independent variables, $\theta$ is the vector of community-level regression coefficients, $\varepsilon_{ij}$ is the *iid* error term of standard logistic distribution with mean 0 and variance $\sigma^2 = \pi^2 / 3$ (Johnson, Kotz, & Balakrishnan, 1995), and the random variable $\nu_j$ is mutually independent of the error term $\varepsilon_{ij}$ with normal distribution of mean zero and between community variance $\tau^2$.

The maximum likelihood method is used to estimate all the unknown coefficients and the cut points $\gamma_1 < \gamma_2 < \cdots < \gamma_5$ to eventually estimate $y_{ij}$ on the basis of the latent variable as

$$y_{ij} = \begin{cases} 0, & \text{if } y_{ij}^* \leq \gamma_1, \\ 1, & \text{if } \gamma_1 < y_{ij}^* \leq \gamma_2, \\ \vdots \\ 5, & \text{if } y_{ij}^* > \gamma_5. \end{cases} \tag{2}$$

From the multilevel mixed-effects ordinal logistic regression model described above, the first five models can be derived under the following conditions:

- Model 5: if the community-level variables $Z_j$ are omitted from the model.
- Model 4: if $\nu_j$ become constant over EAs, so that $\nu_j + \alpha = \alpha_0$ becomes constant.
- Models 1-3: if the ordinal response variable has only two categories, then Models 4-6 become Models 1-3, respectively.

Model 7 is the generalized version of the Model 4. Model 4 requires to satisfy the proportional odds assumption (Long & Freese, 2014), which states that the slopes of each of the independent variables would be identical over every cumulative partition of the response variable. Brant Wald-type test (Brant, 1990) is used in this study to test this assumption. However, the generalized ordered logistic regression model is not restricted to this assumption (Richard, 2006). The mathematical specification of the generalized ordered logistic model for the ordinal response variable (Clogg & Shihadeh, 1994) under the paper is as follows.

$$P(y_i > k \mid X_i) = F(X_i'\beta_k) = \frac{\exp(\alpha_k + X_i'\beta_k)}{1 + \exp(\alpha_k + X_i'\beta_k)}, \quad k = 0, 1, 2, 3, 4. \tag{3}$$

### Performance criteria of the models

To compare the performance of the models under the paper, the Akaike information criteria (AIC)(Akaike, 1974) and Bayesian information criteria (BIC) (Schwarz, 1978) are used. Once we use the multilevel models, we report the likelihood ratio (LR) test statistic (Greene, 2012) along with $p$-value to compare the model with the corresponding ordinary version.

## Results and Discussion

## Evidence from descriptive analysis

### Univariate distribution of the original IPV questions and response variables

The distributions of the women's attitude towards the IPV in terms of the original questions and the generated response variables are presented in Table 1. In general, it reveals that the percentage of women permissive to IPV for each of the situations, except for burning the food, decreases in BDHS 2014 survey compared to that evident from the analysis in Jesmin (Jesmin, 2017) based on BDHS 2011 data. This indicates a slight improvement in forming the right attitude towards IPV.

As an individual case, the highest frequency of permissiveness (about 21%) is evident due to arguing with the husband, followed by neglecting the children (16%), and if she goes out without telling her husband (15%). Only around 8% of the women believe that beating is justified if they sometimes refuse to have sex with their husbands. There is about 95.33% of the women believed that if anyone burns the food unintentionally, she might not be beaten by her husband. Around 29% of the women accept that beating is justified if at least one of the five situations occurs, and only 1.92% of the women believe that beating a wife is justified for each of the five situations. The detail distributions of other combinations are also reported in the Table.

Table 1: Distribution of the Respondents by Response Variables and Original Attitude Responses on IPV

| Variable | Percentage |
|---|---|
| **Original Questions** | |
| Wife beating is justified if she goes out without telling her husband | 15.07 |
| Wife beating is justified if she neglects the children | 15.65 |
| Wife beating is justified if she argues with her husband | 20.79 |
| Wife beating is justified if she refuses to have sex | 7.78 |
| Wife beating is justified if she burns the food | 4.67 |
| **Binary Response Variables** | |
| Permissive to at least one of the five IPV situations | 29.36 |
| **Ordinal Response Variable** | |
| Permissive to none of IPV situations | 70.64 |
| Permissive to any one of the five IPV situations | 11.42 |
| Permissive to any two of the five IPV situations | 8.05 |
| Permissive to any three of the five IPV situations | 5.28 |
| Permissive to any four of the five IPV situations | 2.69 |
| Permissive to all of the five IPV situations | 1.92 |

### *Bivariate analysis of the characteristics by binary response variable*

Table 2 reports the bivariate distribution of the individual and community-level variables against the dichotomous response variables along with the *p*-values generated by using the *t*-test and $\chi^2$-test depending on the scale of measurement of the variables. This distribution would help to get the initial idea about the significant factors on which the differences may occur on the response variables.

The distribution of age of the respondent over the binary response variable indicates that the average age of those permissive to the IPV and their counterpart is around 31 years, which is equivalent to the overall average. The *p*-value clearly demonstrates that there is no difference in having the right or wrong attitude towards the IPV due to the age of the respondents.

Shockingly, the average age at first marriage of the women under the paper is around 16 years, while the global average age is 23.6 years (Hertrich, 2017) and the legal minimum age for marriage is 18 years in Bangladesh (CMRA, 2017). Moreover, the women who approve beating is justified in at least one of the five situations have a significantly lower mean age at first marriage compared to those who do not approve.

In order to see the differences over the educational status of the women, the mean distribution over the binary response variable is presented in Table 3. It is evident that the average years of education of women is only around 5 years, which is very low. Despite having a low overall average, it is evident that higher education significantly lowers the likelihood of the permissiveness to the IPV. A similar pattern of significant differences is observed for mean autonomy or decision score, access to the information index, husband's total years of education, and normalized wealth score. These differences indicate that participation in making major household decisions, having access to the information, higher educational attainment of the husband, and well wealth status ensure to possess the right attitude towards the IPV. Household size is also taken into consideration in this paper, and no difference is found over the response variable.

Table 2: Distribution of Individual and Community Level Covariates by Binary Response

| Variable | Permissive to at Least one IPV Situation | | Total | *p*-value |
|---|---|---|---|---|
| | No | Yes | | |
| **Individual Level Variables** | | | | |
| Age | 30.95 | 31.18 | 31.02 | 0.13 |
| Age at first marriage | 16.05 | 15.40 | 15.86 | 0.00 |
| Respondent's total years of education | 5.67 | 4.40 | 5.30 | 0.00 |
| Any work in last 12 months | | | | |
|     No | 71.20 | 28.80 | 100.00 | 0.02 |
|     Yes | 69.54 | 30.46 | 100.00 | |
| Autonomy/decision score | 1.80 | 1.57 | 1.73 | 0.00 |
| Religion (Muslim) | | | | |
|     Non-Muslim | 74.71 | 25.29 | 100.00 | 0.00 |
|     Muslim | 70.20 | 29.80 | 100.00 | |
| NGO member | | | | |
|     No | 71.85 | 28.15 | 100.00 | 0.00 |
|     Yes | 68.19 | 31.81 | 100.00 | |
| Access to information index | 1.51 | 1.14 | 1.40 | 0.00 |
| Husband's total years of education | 5.90 | 4.48 | 5.48 | 0.00 |
| Household size | 5.38 | 5.41 | 5.38 | 0.35 |
| Place of residence (Rural) | | | | |
|     Urban | 74.93 | 25.07 | 100.00 | 0.00 |
|     Rural | 68.37 | 31.63 | 100.00 | |
| Normalized wealth score | 0.352 | 0.290 | 0.334 | 0.00 |
| Division (Dhaka) | | | | |
|     Other Division | 69.47 | 30.53 | 100.00 | 0.00 |
|     Dhaka Division | 76.24 | 23.76 | 100.00 | |
| **Community Level Variables** | | | | |
| Proportion of poorest HHs | 0.17 | 0.21 | 0.18 | 0.00 |
| Proportion of illiterate | 0.23 | 0.25 | 0.24 | 0.00 |
| Autonomy/decision score | 1.76 | 1.66 | 1.73 | 0.00 |

*Note: For continuous control variables, p-values are obtained from the t-test for mean comparison over the response variable, while, for dichotomous control variables, p-values are obtained from the chi-square test for measuring association.*

The distribution of categorical variables over the response variable includes whether the women worked in the last 12 months preceding the survey, the religious status of either Muslim versus Non-Muslim, whether the member of NGO or non-member, the place of residence either in rural or urban areas and living in Dhaka division or other divisions. The chi-square test has been carried out to see whether there is any difference due to each of the categorical characteristics of the women. The distribution of the working status reveals that more women had justified IPV if they were involved in work compared to their counterparts, and this difference is significant at 5% level of significance. Among the Muslim women, around 30% responded in favor of beating if at least one of the IPV situations occurs compared to around that of 25% who believe in other religious faith. Similarly, significantly higher likelihood in favor of justifying beating among NGO members and those who are living in rural areas is evident compared to their respective counterparts'. On the other hand, a significantly lower percentage is evident among those living in the Dhaka division than that of other divisions of the country.

Three community-level variables are also found significantly different over the response variable. Expectedly, the group who justifies the IPV, the higher proportion of poorest households, and illiterate women are found in the respective communities. On the other hand, a significantly lower average community level of decision score is evident for the women who are tolerant of IPV compared to their counterparts.

**Evidence from statistical modeling**

***Permissiveness to at least one IPV situation***

Table 3 summarizes the results extracted by using the ordinary (Model 1) as well as the mixed-effects multilevel logistic regression models (Models 2-3). Overall, the signs of the coefficients and the corresponding significances are consistent in all models. By the *p*-values, three different sets of significant independent variables have been identified. The first set includes age at first marriage, respondent's total years of education, autonomy or decision score, access to information index, husband's total years of education, normalized wealth score, and division dummy. All these variables have a significant effect on the permissiveness to at least one of the IPV situations. Other than the division dummy, the higher the value of the variables, the lower the likelihood to justify the IPV. The coefficient of division dummy shows that the likelihood is significantly lower if the women living in the Dhaka division compared to other divisions.

The second set of significant variables include the religion of the women and whether they are a member of NGO or not. It unveils that Muslim women are more tolerant of IPV compared to women from other religions. There is a longstanding belief that microfinance services have a positive impact on women's empowerment (Weber & Ahmed, 2014; Amin, Becker & Bayes, 1998) and other awareness. Surprisingly, it reflects from our analysis that NGO members are more positive to justify IPV situations compared to non-members.

Table 3: Logistic Regression and Mixed-effects Multilevel logistic regression Models of Permissiveness to at Least One of the IPV Situations

| Variable | Logistic Coef. (*p*-value) | Multilevel Logistic Coef. (*p*-value) | Multilevel Logistic Coef. (*p*-value) |
|---|---|---|---|
| **Individual Level Variables** | | | |
| Age | -0.002 (0.281) | -0.002 (0.486) | -0.002 (0.421) |
| Age at first marriage | -0.032 (0.000) | -0.036 (0.000) | -0.035 (0.000) |
| Respondent's total years of education | -0.032 (0.000) | -0.036 (0.000) | -0.037 (0.000) |
| Any work in last 12 months | -0.002 (0.966) | 0.020 (0.621) | 0.018 (0.645) |
| Autonomy/decision score | -0.117 (0.000) | -0.107 (0.000) | -0.095 (0.000) |
| Religion (Muslim) | 0.227 (0.000) | 0.219 (0.004) | 0.201 (0.008) |
| NGO member | 0.070 (0.058) | 0.102 (0.010) | 0.104 (0.009) |
| Access to information index | -0.066 (0.000) | -0.056 (0.004) | -0.052 (0.007) |
| Husband's total years of education | -0.014 (0.008) | -0.013 (0.021) | -0.013 (0.017) |
| Household size | 0.008 (0.232) | 0.003 (0.714) | 0.001 (0.926) |
| Place of residence (Rural) | -0.009 (0.818) | 0.027 (0.702) | -0.019 (0.789) |
| Normalized wealth score | -0.887 (0.000) | -0.905 (0.000) | -0.837 (0.000) |
| Division (Dhaka) | -0.223 (0.000) | -0.267 (0.002) | -0.172 (0.044) |
| **Community Level Variables** | | | |
| Proportion of poorest HHs | | | 0.212 (0.285) |
| Proportion of illiterate | | | -0.434 (0.100) |
| Mean autonomy/decision score | | | -0.397 (0.000) |
| Constant | 0.264 (0.075) | 0.214 (0.214) | 0.977 (0.000) |
| **Random effects** | | | |
| Between community variance $\tau^2$ (SE) | | 0.387 (0.036) | 0.365 (0.035) |
| Intraclass correlation $\rho$ (SE) | | 0.105 (0.009) | 0.100 (0.008) |
| **Model Parameters** | | | |
| Log likelihood | -10451.74 | -10193.65 | -10180.45 |
| AIC | 20931.48 | 20417.30 | 20396.90 |
| BIC | 21040.55 | 20534.15 | 20537.12 |
| Likelihood Ratio $\chi^2$ (*p*-value) | | 516.19 (0.000) | 475.99 (0.000) |

In the third set, we list the variables which do not have any significant effect in determining the women's attitude towards the violence. These variables include the age of the women at survey time, whether they involve in any work in the last 12 months, household size, and living either in rural or urban areas. Despite insignificant effects, this paper listed these variables to avoid the paradox of having a significant contribution in controlling the attitude of the women.

This paper considers three community-level variables, including the proportion of women illiterate, the proportion living in the poorest household, and the mean autonomy or decision score at the community level. The estimated mixed-effects multilevel model unveils that though both the individual level of women's education and household wealth index have significant effects on the women's attitude toward IPV, however, the corresponding community level does not have a significant effect. On the other hand, the mean autonomy or decision score over the community level still has a significant impact on reducing the likelihood of permissiveness towards IPV.

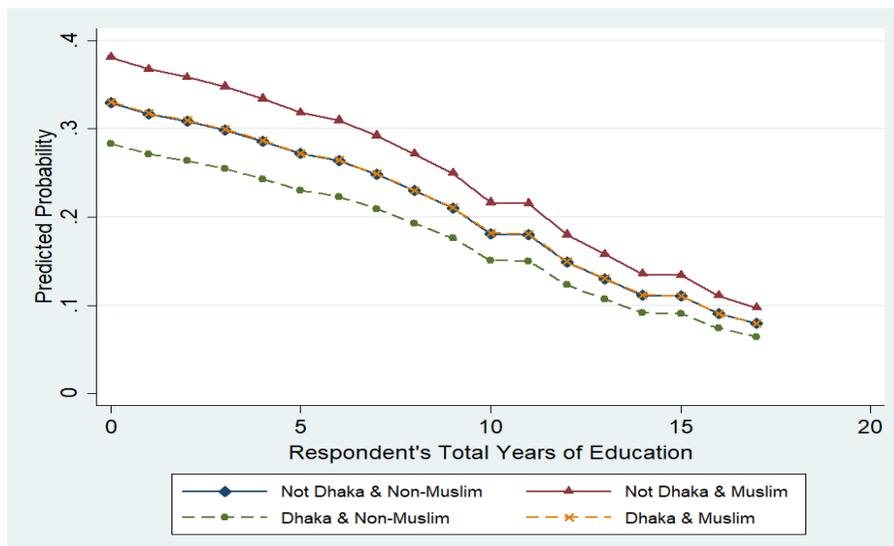

Figure 2: Average Predicted Probabilities by Division, Religion against the Age at First Marriage

The trends of the predicted probabilities obtained from the binary logistic regression over the combination of the predictors are analyzed by graphical representation. Figure 2 exhibits the predicted probabilities against the age at first marriage of the women by division and religion when all other variables are controlled. This figure facilitates to make several conclusions regarding the variables. Firstly, the uniform highest permissiveness is evident for Muslim women and those living in other than the Dhaka division compared to their respective counterparts. On the other hand, the lowest is apparent for those Non-Muslim and living in the Dhaka division. The predicted probabilities of the combination Not Dhaka and Non-Muslim is pulled up to the middle from the lowest category because of including the respondents from other divisions and similarly, the category Not Dhaka and Muslim with the highest permissiveness is pulled down to the middle because of replacing the respondents of other divisions by the respondents of Dhaka divisions.

Secondly, regardless of the city of residence or religion, the predicted probabilities show a downward trend as the increase of age at first marriage. This indicates higher the age at first marriage lower the tendency to justify IPV wrongly. Thirdly, if the legal age at marriage of 18 years is maintained, then it results in substantially lower permissiveness compared to that observed in general (around 29%, see Table 1). Remarkably, it is lower than 20% if age at marriage is just higher than 21 years and even goes to below 10% at some point around 28 years.

Similar figures can be constructed by keeping the religion and division indicator as they are, but by replacing the age at 1st marriage, for example, by respondent's total years of education, autonomy or decision score, access to information index, and husband's total years of education respectively.

*Permissiveness to the ordinal IPV index*

Table 4 reports the estimated coefficients along with corresponding *p*-values and the estimates of other parameters of Models 4-6. It reveals that the signs of the coefficients along with the *p*-values of all the characteristics, but the community proportion of illiterate, are found consistent with that obtained from the models with the binary response variable. It shows that the community proportion of women illiterate has a significant effect in reducing the positive attitude towards IPV at 10% level of significance, which is consistent with the previous study based on BDHS 2011 data (Jesmin, 2017).

Table 4: Ordinal Logistic and Mixed-effect Multilevel Ordinal Logistic Model of Permissiveness to IPV

| Variable | Ordinal Logistic Coef. (*p*-value) | Multilevel Ordinal Logistic Coef. (*p*-value) | Multilevel Ordinal Logistic Coef. (*p*-value) |
|---|---|---|---|
| **Individual Level Variables** | | | |
| Age | -0.002 (0.364) | -0.001 (0.619) | -0.001 (0.539) |
| Age at first marriage | -0.032 (0.000) | -0.035 (0.000) | -0.035 (0.000) |
| Respondent's total years of education | -0.035 (0.000) | -0.040 (0.000) | -0.042 (0.000) |
| Any work in last 12 months | 0.008 (0.827) | 0.036 (0.358) | 0.035 (0.375) |
| Autonomy/decision score | -0.126 (0.000) | -0.116 (0.000) | -0.105 (0.000) |
| Religion (Muslim) | 0.256 (0.000) | 0.247 (0.001) | 0.229 (0.002) |
| NGO member | 0.052 (0.146) | 0.080 (0.037) | 0.082 (0.033) |
| Access to information index | -0.074 (0.000) | -0.065 (0.000) | -0.062 (0.001) |
| Husband's total years of education | -0.012 (0.017) | -0.012 (0.031) | -0.012 (0.025) |
| Household size | 0.005 (0.424) | -0.001 (0.886) | -0.003 (0.689) |
| Place of residence (Rural) | -0.016 (0.691) | 0.024 (0.735) | -0.026 (0.719) |
| Normalized wealth score | -0.957 (0.000) | -0.980 (0.000) | -0.909 (0.000) |
| Division (Dhaka) | -0.222 (0.000) | -0.265 (0.002) | -0.165 (0.058) |
| **Community Level Variables** | | | |
| Proportion of poorest HHs | | | 0.248 (0.218) |
| Proportion of illiterate | | | -0.466 (0.083) |
| Mean autonomy/decision score | | | -0.406 (0.000) |
| /cut1 | -0.303 | -0.256 (0.130) | -1.036 (0.000) |
| /cut2 | 0.361 | 0.452 (0.008) | -0.328 (0.166) |
| /cut3 | 1.067 | 1.194 (0.000) | 0.414 (0.081) |
| /cut4 | 1.901 | 2.055 (0.000) | 1.275 (0.000) |
| /cut5 | 2.811 | 2.981 (0.000) | 2.201 (0.000) |
| **Random effects** | | | |
| Between community variance $\tau^2$ (SE) | | 0.417 (0.037) | 0.393 (0.036) |
| Intraclass correlation $\rho$ (SE) | | 0.113 | 0.107 |
| **Model Parameters** | | | |
| Log likelihood | -17892.91 | -17582.76 | -17569.16 |
| AIC | 35821.82 | 35203.53 | 35182.32 |
| BIC | 35962.05 | 35351.55 | 35353.70 |
| Likelihood Ratio $\chi^2$ (*p*-value) | | 620.29 (0.000) | 574.66 (0.000) |

Regarding the proportion of women living in the poorest household in the community, the previous study reported that higher the proportion lowers the likelihood to justify the IPV, which is inconsistent with the literature (Jesmin, 2017). In our paper, although it is not evident as significant, however, the sign is consistent with that shown as individual-level effect as well as in the literature.

Moreover, the place of residence in either rural or urban areas does not affect the women's attitude significantly. Whereas, this paper shows that living in the Dhaka division, which is more developed compared to other divisions, has a significant effect in reducing the permissiveness to IPV. By considering the facilities in the Dhaka division compared to other divisions, the lower likelihood of permissiveness to IPV is justified. However, for the same indicator, the result was opposite that reported in Jesmin (Jesmin2017). A further survey would be helpful to conclude which direction is more likely with certainty.

The Brant test results to check the proportional odds assumption are reported in Table 5. The *p*-value of the overall test indicates the violation of the proportional odds assumption of the ordinal logistic regression model. From the tests of individual-level independent variables, the violation has been evident for the respondent's total years of education at 5% level of significance and normalized wealth index at 1% level of significance. This implies that the slopes of education and wealth index are significantly different across the categories of the ordinal response variable.

Table 5: Brunt Test of Proportional Odds Assumption

| Test | $\chi^2$ | *p*-value | df |
|---|---|---|---|
| **Overall** | 132.03 | 0.000 | 52 |
| **Individual-Level Variables** | | | |
| Age | 2.76 | 0.599 | 4 |
| Age at first marriage | 5.24 | 0.264 | 4 |
| Respondent's total years of education | 12.33 | 0.015 | 4 |
| Any work in last 12 months | 2.85 | 0.584 | 4 |
| Autonomy/decision score | 9.00 | 0.061 | 4 |
| Religion (Muslim) | 6.06 | 0.195 | 4 |
| NGO member | 5.96 | 0.202 | 4 |
| Access to information index | 6.50 | 0.165 | 4 |
| Husband's total years of education | 5.42 | 0.247 | 4 |
| Household size | 9.33 | 0.053 | 4 |
| Place of residence (Rural) | 3.47 | 0.482 | 4 |
| Normalized wealth score | 16.25 | 0.003 | 4 |
| Division (Dhaka) | 5.67 | 0.225 | 4 |

***Permissiveness to the IPV index: generalized ordered logistic***

The ordinal response variable has a total of 6 categories having the values from 0 to 5. Thus, estimating the generalized ordered logistic regression model yields the estimates of 5 different sets of coefficients, including intercept corresponding to the cumulative responses. The response groups are defined as (i) $y > 0$ versus $y \leq 0$, (ii) $y > 1$ versus $y \leq 1$, (iii) $y > 2$ versus $y \leq 2$, (iv) $y > 3$ versus $y \leq 3$, and (v) $y > 4$ versus $y \leq 4$. The estimated coefficients are presented in Table 6.

Table 6: Generalized Ordered Logistic Model of Permissiveness to IPV

| Variable | y>0 coef. (*p*-value) | y>1 coef. (*p*-value) | y>2 coef. (*p*-value) | y>3 coef. (*p*-value) | y>4 coef. (*p*-value) |
|---|---|---|---|---|---|
| **Individual Level Variables** | | | | | |
| Age | -0.003 (0.222) | 0.000 (0.900) | 0.000 (0.907) | 0.000 (0.947) | -0.002 (0.774) |
| Age at first marriage | -0.032 (0.000) | -0.031 (0.000) | -0.044 (0.000) | -0.046 (0.003) | -0.026 (0.295) |
| Respondent's total years of education | -0.032 (0.000) | -0.039 (0.000) | -0.052 (0.000) | -0.079 (0.000) | -0.113 (0.000) |
| Any work in last 12 months | -0.001 (0.988) | 0.050 (0.239) | 0.028 (0.606) | -0.011 (0.883) | 0.067 (0.564) |
| Autonomy/decision score | -0.117 (0.000) | -0.140 (0.000) | -0.165 (0.000) | -0.175 (0.000) | -0.225 (0.000) |
| Religion (Muslim) | 0.227 (0.000) | 0.346 (0.000) | 0.420 (0.000) | 0.431 (0.003) | 0.521 (0.023) |
| NGO member | 0.070 (0.057) | 0.017 (0.694) | -0.006 (0.918) | -0.071 (0.356) | 0.024 (0.840) |
| Access to information index | -0.065 (0.000) | -0.095 (0.000) | -0.120 (0.000) | -0.133 (0.001) | -0.151 (0.020) |
| Husband's total years of education | -0.014 (0.000) | -0.013 (0.042) | 0.000 (0.994) | 0.001 (0.909) | -0.001 (0.966) |
| Household size | 0.008 (0.257) | 0.003 (0.710) | -0.011 (0.271) | -0.031 (0.045) | 0.002 (0.945) |
| Place of residence (Rural) | -0.008 (0.837) | -0.030 (0.538) | -0.033 (0.595) | -0.046 (0.601) | 0.129 (0.349) |
| Normalized wealth score | -0.880 (0.000) | -1.183 (0.000) | -1.372 (0.000) | -2.126 (0.000) | -1.816 (0.002) |
| Division (Dhaka) | -0.228 (0.000) | -0.237 (0.000) | -0.141 (0.056) | -0.098 (0.353) | 0.088 (0.566) |
| Constant | 0.269 (0.070) | -0.364 (0.043) | -0.739 (0.002) | -1.099 (0.001) | -2.631 (0.000) |
| **Model Parameters** | | | | | |
| Log likelihood | -17818.07 | | | | |
| AIC | 35776.15 | | | | |
| BIC | 36321.47 | | | | |

It is evident from the first model that there are several variables that have a significant effect on the attitude of the women. In comparing to the other models presented above, we have the same set of significant predictors, including age at first marriage, women's total years of education, autonomy or decision score, religion, access to information, husband's total years of education, normalized wealth score and division indicator. They all are significant at 1% level of significance. At 10% level of significance, the NGO membership has a significant effect, and the coefficient indicates that the NGO members justify IPV situations more likely than their counterparts.

In the second model, women who justify more than one IPV situations are compared with the reference of justifying at least one IPV situation. There are no changes in the significant predictor list compared to the first model; however, slight changes are observed in the coefficients. The changes in the coefficients of the predictors except age at first marriage and husband's education make the difference in possessing the justifying attitude of IPV wider between the response groups. For instance, the respondent with more years of education is more likely to belong in the reference group than the group justifying more than one IPV situation. Similar effects are observed for decision score, access to information index, normalized wealth score, and division indicator. The coefficient of religion indicates that compared to the first model, Muslim women are more likely to belong in the cumulative group rather than the reference group. Thus, it implies that Muslim women are more tolerant of more than one IPV situations than Non-Muslim. The similar conclusion of higher the age at first marriage of women and higher the level of education of husband lowers the likelihood of justifying the IPV can be taken from the second model, but the differences in the magnitude of coefficients compared to the first model are not considerable.

A similar pattern on the changes of most of the coefficients is evident for the third model. However, the coefficient of the predictor husband's education becomes 0, while the division indicator is found insignificant at 5% level of significance. The fourth model compares the respondents who justify more than three of the five IPV situations, with the respondents justifying at least three IPV situations. The sign of the coefficients of this model is similar to that of previous models. Only the differences are evident in their magnitude. Also, household size is found significant at 5% level of significance for the first time, and larger family size indicates lowering the permissiveness.

In the final model, age at first marriage and division indicator are found insignificant, and the rest of the independent variables which are found significant from all the previous models are also found significant here. This indicates that age at first marriage and division indicators are not enough to significantly differentiate the respondents between the higher IPV justification group and its counterpart. On the other hand, significant variables, including respondent's total years of education, decision score, access to information index, and wealth index, have significant effects on women not to justify all of the five IPV situations. However, religion still determines Muslim women to belong to the highest justifying group.

### *Assessment of the models*

The likelihood ratio test is used to check the performance of the mixed-effect multilevel models or, more specifically, random-intercept models for both binary and ordinal response variables over the respective ordinary models. Tables 3 and 4 reported the likelihood ratio statistics and the corresponding *p*-values which indicate a significant improvement. The improvement is also evident from the presented AIC and BIC statistics. However, no substantial variations are apparent in these statistics over the multilevel models with and without the community-level variables. In the generalized ordered logistic model, the AIC and BIC are found slightly higher than the ordinal and mixed-effects multilevel ordinal models but meets the objective to identify the significant determinants.

**Conclusions and Recommendations**

**Conclusions**

Intimate partner violence (IPV) is a deep-rooted problem for society worldwide, and still, the prevalence rate is very high. Surprisingly, the victim women, to some extent, justify the violence against themselves by their partners with wrong self-perceptions. This paper uses a nationally representative BDHS 2014 data to explore the important determinants of the women's attitude towards five IPV situations. The responses of 17,863 women are analyzed to meet the objectives of the paper. For appropriate identification of the significant determinants, a series of individual-level and community-level variables are investigated by two response variables, which are constructed to measure the extent at which the women justify the IPV. Along with the univariate and bivariate distributions, seven regression models are used to estimate the effects of the predictors on two response variables.

The univariate distribution of the women's responses on the five original questions exhibits that ranges from 4.67% to 20.79% of women believe that beating by the husband is justified. The highest percentage is reported corresponding to the reason for arguing with the husband, while the lowest is due to burning foods. Though it is about 3% lower than the result from Jesmin (Jesmin, 2017), the estimated 29% of women justifying the beating for at least one of the five situations is very alarming. It is also found that around 2% of women believe that beating is justified in all of the five IPV situations.

The bivariate distribution and statistical models reveal the set of the important determinants including age at first marriage, respondent's education, decision score, religion, NGO membership, access to information, husband's education, normalized wealth score, and division indicator have significant effects on the women's attitude towards IPV. It is evident that other than religion, NGO membership, and division indicator, the higher the value of the variable, the lower the likelihood of justifying IPV. However, being a Muslim, NGO member, and resident of other divisions, women are found more tolerant of IPV from their respective counterparts. Among the three community-level variables, only the mean decision score is found significant in lowering the likelihood.

**Recommendations**

- This paper shows the shocking average age at first marriage despite having the legal age of marriage. As higher the age at first marriage ensures the likelihood of possessing the right attitude to IPV, the government, private organizations, international donors, and other stakeholders should work to ensure at least the legal age at first marriage.
- The educational qualifications of both respondents and their husbands have a positive impact of having a righteous attitude towards IPV. Ensuring education for all in this regard would work better for the current as well as the future generation.

- The participation of women in the household's major decision-making process helps them to justify IPV situations appropriately. The initiatives from household to secure women's participation is necessary.
- Referring to the conclusions, Muslim women, compared to women of other religions, are found more tolerant of IPV situations. It is possible that the Muslim community is not completely aware of this tendency. To make this public and motivate the community by the religious leaders would be a solution to divert the wrong attitude.
- Access to information has a significant effect on lowering the likelihood of wrongly justifying IPV situations. Thus, facilitating well access to information through TV, radio, and newspapers is effective in creating awareness regarding IPV situations.
- The government should take some initiatives for the disadvantaged group to maintain the minimum standard of economic activity. This would facilitate the group to have the right attitude towards IPV.
- The division indicator shows that the women living in the Dhaka division are more likely to have the right attitude towards IPV than that of other divisions. It is necessary to ensure all the facilities in other divisions similar to the Dhaka division.

This study suggests working all the stakeholders on these recommendations based on the significant determinants to ensure to possess the right attitudes towards the IPV, and thus help to take away this deep-rooted problem from society.

**Declarations:**

***Ethics approval and consent to participate:*** Not applicable. The used dataset does not require IRB approval.

***Consent for publication:*** Not applicable.

***Availability of data and materials:*** Data available on request from the site https://dhsprogram.com/data/available-datasets.cfm.

***Competing interests:*** The authors declare that they have no competing interests.

***Funding:*** The authors do not receive any financial support for the research, authorship, and/or publication of this article.

***Author's Contributions:*** MTFK analyzed the data and prepared the manuscript under the direct supervision of LQ. In every step of preparing and finalizing the manuscript, both authors have the same contribution.

***Acknowledgment:*** The authors acknowledge the Demographic and Health Survey (DHS) Program authority for providing access to the Bangladesh Demographic and Health Survey Data 2014 for this study.